\documentclass[twocolumn,prl,floatfix]{revtex4}
\usepackage{graphicx}
\usepackage{epstopdf}
\usepackage{rotating}
\usepackage{amsmath}
\usepackage{amsfonts}
\usepackage{amssymb}
\usepackage{enumerate}
\usepackage{longtable}
\usepackage{dcolumn}
\usepackage{bm}
\usepackage{xcolor}

\begin{document}

\title{Robustness of entropy plateaus: A case study of triangular Ising antiferromagnets}

\author{Owen Bradley, Chunhan Feng, Richard T. Scalettar and Rajiv R. P. Singh}
\affiliation{Department of Physics, University of California Davis, CA 95616, USA}

\date{\rm\today}

\begin{abstract}
Residual entropy is a key feature associated with emergence in many-body systems. From 
a variety of frustrated magnets to the onset of
spin-charge separation in Hubbard models and fermion-$Z_2$-flux variables in Kitaev models, the freezing
of one set of degrees of freedom and the establishment of local constraints are marked by a plateau in entropy
as a function of temperature. Yet,
with the exception of the rare-earth pyrochlore family of spin-ice materials, evidence for such plateaus
is rarely seen in real materials, raising questions about their robustness.
Following recent experimental findings of the absence of such plateaus in the
triangular-lattice Ising antiferromagnet (TIAF) TmMgGaO$_4$ by Li et al, we 
explore in detail the existence and rounding of entropy plateaus in TIAF.
We use a transfer matrix method 
to numerically calculate the properties of the system at different temperatures and magnetic fields,
with further neighbor interactions and disorder.
We find that temperature windows of entropy plateaus exist
only when second-neighbor interactions are no more than a couple of percent of the nearest-neighbor ones,
and they are also easily destroyed by disorder in the nearest-neighbor exchange variable, thereby explaining the challenge in observing such effects. 
\end{abstract}

\pacs{}

\maketitle

\section{I. Introduction}

Residual entropy is a hallmark of frustrated systems, reflecting the emergence of local constraints
or new degrees of freedom distinct from the microscopic ones \cite{balents}. One of the earliest theoretical works
in this direction was the calculation of residual entropy associated with the establishment of ice
rules in water done by Pauling \cite{pauling}. It is now well established that such a strongly constrained
phase has an analog in magnetic systems known as spin-ice \cite{gingras,moessner,ice-review}.
Such a classical spin-liquid exhibits
residual entropy \cite{ramirez, ke, cornelius} and supports magnetic-monopole excitations. Quantum fluctuations in such a system can lead to a 
highly resonating quantum spin-liquid phase
with emergent quantum electrodynamics.

In models of geometrically frustrated magnets, such residual entropy is widespread \cite{mila}. But, how robust are they in
real materials? In fact, the issue is much broader than geometric frustration. In recent years, there has been a lot of interest in Kitaev materials \cite{khaliullin,hykee,valenti,motome,trebst}.
At a microscopic level, the honeycomb-lattice Kitaev model describes spins interacting
with anisotropic bond-direction dependent exchange interactions \cite{kitaev}. Yet, the model can be exactly mapped onto
one of Majorana fermions and $Z_2$-valued fluxes. As the fermions reach their degeneracy temperature
or freeze-out if they have a gapped spectrum, an entropy plateau sets in \cite{nasu}. Indeed, some hints of
entropy plateaus have been seen in experiments \cite{kitaev-exp,kitaev-exp2}. The plateaus are far more robust than the soluble
models. For example, spin-$S$ models show even more interesting possibilities of entropy plateaus \cite{bss,crd,ktn,oitmaa}.
When Kitaev couplings are the same along all three axes, there are incipient plateaus at an entropy of $\ln{(2S+1)}/2$, 
which keeps increasing with spin $S$. The physical mechanism behind such large entropy values at
the plateau with increasing $S$ is not well understood.

Entropy plateaus are also a prominent feature of correlated electron
Hamiltonians such as the Hubbard and periodic Anderson models \cite{deleo11,li14,paiva}.
In the regime of strong on-site interaction $U$, there are clearly
distinct charge and spin energy scales.  At the higher scale,
$T \sim U$, a drop in entropy occurs when doubly occupied sites
are frozen out.  At the lower scale, $T \sim J = 4t^2/U$, a second
drop occurs that is associated with the development of antiferromagnetic (AF)
correlations.
Studies of the specific heat $C(T)$ in one
dimension \cite{jutner98,shiba72,schulte96},
and in infinite dimensions, i.e.~within dynamical mean field theory
(DMFT) \cite{georges93,vollhardt97,chandra99},
suggested the disappearance of an entropy plateau as $U \rightarrow t$,
and hence the two energy scales merge.
However, quantum Monte Carlo calculations for the
half-filled Hubbard model on a two-dimensional square lattice
revealed that the
two peak structure in $C(T)$ is robust, surviving even down
to $U \sim t$.  An interesting feature of this robustness was an
apparent interchange in the ``driving force'' of the entropy reduction.
At strong $U$, changes in the {\it potential energy} led to the high-$T$
specific heat peak, while at small $U$, it is the changes in the {\it kinetic
energy} that yields the peak at higher temperature.

Preservation of the entropy plateaus, in these systems, appears to be linked to
the AF order.  On a honeycomb lattice \cite{tang13}, the
two peaks in $C(T)$ merge as $U$ is reduced, with a resultant
destruction of the plateau.  The most natural
explanation is that, unlike the
square lattice where antiferromagnetism exists down to $U=0$, the honeycomb lattice
has a quantum critical point $U_c/t \sim 4$, below which antiferromagnetism disappears \cite{paiva, sorella}.

The Ising antiferromagnet on the triangular lattice is an iconic problem
in frustrated magnetism where an exact residual ground state entropy was
first calculated by Wannier \cite{wannier1950,wannier1973}. 
Several materials, including
CeCd$_3$As$_3$ \cite{liu}, FeI$_2$ \cite{katsumata}, and TmMgGaO$_4$
\cite{cevallos, li}, have been identified experimentally as triangular-lattice Ising antiferromagnet (TIAF) systems
owing to the strong Ising nature of their constituent spins. 
Despite being of such central interest,
there are few (or no) experimental systems where such residual entropy
has been observed. 
Very recently, Li et al \cite{li} investigated the triangular-lattice Ising
antiferromagnetic material TmMgGaO$_4$.  They measured the heat capacity
and entropy of the system as well as the magnetization as a function of
an applied magnetic field. Li et al found a complete absence of entropy plateaus and rounded magnetization plateaus, with roundings that are only partly thermal and partly reflect the presence of quenched impurities.  

The TIAF has been studied over the years using a variety of analytical and numerical methods. Such studies have
determined the phase diagram, minimum energy spin configurations, as
well as entropy and specific heat curves for finite size clusters 
\cite{metcalf1973, kinzel, metcalf1974, metcalf1978, hwang, rastelli}. Several numerical studies of disordered TIAF (and ferromagnetic) systems have also been performed,
including investigations of random site vacancies, diluted lattices,
varying bond lengths, and disorder in the applied field \cite{gu,
kurbah, zukovic}.

The purpose of this paper is to explore the rounding or absence of
entropy and magnetization plateaus in the TIAF as a function of applied
field due to further neighbor interactions and disorder.  How large a perturbation can the system tolerate before the plateaus disappear altogether? We use a numerical transfer matrix based approach to calculate the thermodynamic
properties. We first confirm that, in the absence of second neighbor
interactions, the magnetization of the pure TIAF jumps from $0$ to
$1/3$ in an infinitesimal field, and then at a field of $B=6$ it jumps again
to full saturation value. The transition field $B=6$ also has a finite
ground-state entropy. 

We next consider antiferromagnetic second-neighbor interactions, as
appropriate for the TmMgGaO$_4$ material. This interaction is shown to
lead to a striped ground-state phase and a finite-temperature phase
transition.  The entropy plateaus are lost rapidly with a fairly small second-neighbor interaction of only a few percent. They are replaced by sharp drops in the entropy at a first-order transition. This is in contrast to the spin-ice system, where the entropy plateaus are very robust and survive even with long-range dipolar interactions \cite{gingras,moessner,ice-review}
and quantum fluctuations \cite{applegate}. If we consider only the nearest-neighbor TIAF with disorder in the exchange interactions, rounded entropy plateaus are quickly destroyed.

When the second-neighbor interaction is about $10\%$
of the nearest-neighbor value, there are magnetization plateaus at
values $0$, $1/3$, $1/2$, and $1$.  
Thermal rounding of the magnetization plateaus is
very gradual.  Despite the finite temperature, the plateaus remain extremely flat, reflecting the energy gap in the system.  The rounding is much stronger with quenched disorder. We find that strong disorder is needed to obtain results that look quantitatively like the experiments, with both plateaus at magnetizations of $1/3$ and $1/2$ becoming significantly rounded.

The plan of the paper is as follows: First, an overview of the model and the
numerical methods is given. We then present entropy $S(T)$ and
specific heat $C(T)$ results for the TIAF system with no magnetic field
present, in the absence of any disorder, for various strengths of the
second nearest-neighbor interaction $J_2$. Disorder in the nearest-neighbor interaction $J_1$ is then introduced and we study its influence on entropy plateaus in the TIAF. We then show the influence of an
applied magnetic field on the form of $S(T)$ and $C(T)$, and we present
magnetization curves for various temperatures and $J_2$ values. Our final
set of results shows the effect of quenched disorder in both $J_1$ and $J_2$ on the magnetization plateaus observed in the TIAF. Two disorder types---box and Gaussian---are compared. We
finally present our conclusions.

\section{II. Model and Methods}

We study a triangular lattice of Ising spins. Both nearest-neighbor (NN) and next-nearest-neighbor (NNN) interactions are considered in an applied magnetic field $B$, perpendicular to the plane of the lattice. The Hamiltonian studied is thus given by,
\begin{equation}
H = -J_1 \sum_{\langle i,j \rangle} S_i S_j - 
J_2 \sum_{\langle \langle i,j \rangle \rangle} S_i S_j - B \sum_i S_i, 
\label{eq:Ham}
\end{equation}
where $J_1$ and $J_2$ denote the NN and NNN coupling strengths, respectively,
and $S_i = \pm 1$ is the Ising spin at site $i$ of the lattice, which
may be aligned parallel or anti-parallel to the applied field. The first
sum is taken over all pairs of NN sites, and the second is a sum over
all NNN pairs. Negative values of $J_1$ and $J_2$ correspond to
antiferromagnetic interactions.

We employ a transfer matrix approach to obtain values of the Helmholtz
free energy $F(T,B)$ for our TIAF system, which is found from the
largest eigenvalue of a suitably constructed transfer matrix. 
We consider a long cylinder-geometry for our calculations,
which implies periodic boundary conditions in the short direction. The
second neighbor interactions demand that the transfer matrix involve two
rows of spins at a time. This is no longer an analytically soluble
problem.  It also limits the sizes of systems that can be studied.
Furthermore, in order for the system to have compatibility with a
three-sublattice structure of the triangular-lattice and for our results not to be artificially affected by the periodic boundary conditions in the
short direction, we need to have a multiple of three spins in each row.
Our results are all based on six spins in a row which requires a
$2^{12}\times 2^{12}$ transfer matrix. We believe that these results
should be reasonably close to the thermodynamic limit, except  near
phase transitions or points where the correlation length becomes large. For the
nearest-neighbor Ising model in zero field, the calculated entropy
curves are close, though clearly not identical, to the exact answer. 

Since the TIAF with nearest and second-neighbor interactions shows a first-order phase transition over a range of parameters, with a jump in the entropy of the system \cite{rastelli},
there will be large finite-size effects near the transition. In a finite system, all thermodynamic
functions must be analytic and hence no jump in entropy is possible. Instead, one would have a
rounded $\delta$-function in the heat capacity per site, whose peak for an $L\times L$ system scales as $L^2$
and peak-width scales as $1/L^2$. In the thermodynamic limit, this becomes a $\delta$-function, whose
integral gives a jump in the entropy, per site, at the transition.

In an $L\times \infty$ transfer-matrix calculation also the largest eigenvalue of the transfer matrix
must be analytic at any finite temperature and hence there can be no jump in a thermodynamic property. 
The correlation length in the infinite direction must be finite (set by $L$) and the jump in entropy must be rounded over
a range of temperatures near the transition. Since the linear dimension $L=6$ of our study is much smaller than
those studied by Monte Carlo simulations of Rastelli et al \cite{rastelli}, the rounding must be
over a wider temperature range. However, we would not expect the transition temperature to be
strongly size dependent for a first-order transition, and our conclusions regarding rounding of entropy 
plateaus at low temperatures should not be affected by this behavior near the transition. 
Comparing our data with those of Rastelli et al will allow us to quantify this effect.

Previous Monte Carlo data \cite{rastelli} are only available for a magnitude of $J_2$
greater than or equal to $0.1$, which pushes the first order transition temperature outside
the plateau region of the nearest-neighbor model.
To accurately evaluate the jump in entropy $\Delta S$ for smaller $J_2$, 
we have performed further Monte Carlo simulations on up to
$96 \times 96$ lattices. At small $J_2$, the transition temperature is very low.
The transition is strongly first order, with clear evidence for
hysteresis.  The internal energy jumps at the transition and $\Delta E$ are
easily read off from the simulations, as is the transition temperature
where the sharp change in energy occurs.  At the transition, we know
that the two states must have equal free energy. Thus we can get the
entropy jump by using the relation $\Delta S = \Delta E/ T_c$. These will also be compared with
the transfer matrix calculations.

Free energies per site are found for a range of temperatures (at fixed
$B$), and over a range of magnetic field values at fixed temperature, from
which the thermodynamic properties $S(T), C(T)$, and $M(B)$ can be computed
easily by taking suitable derivatives. For a triangular lattice $N$
sites wide, with $2P$ rows of spins (i.e.~there are $P$ distinct
`blocks' of two rows, each $N$ sites in width), the partition function
is given by $Z = \sum_{{s_i}}e^{-\beta H}$ where $H$ is as given in
Eq.~(\ref{eq:Ham})
and we take $k_B$ equal to unity, with the sum taken over all spin
configurations. The method relies on the fact that a careful
construction of a particular $2^{2N} \times 2^{2N}$ matrix $M$ allows
one to write the partition function as $Z = \sum_{S_A} M^P(S_A;S_A) =
Tr[M^P]$, where $S_A$ is shorthand for a particular configuration of 2$N$
spins within a block. The partition function is thus given by
\begin{equation}
Z = \lambda_1^P + \lambda_2^P + \lambda_3^P\ldots,
\label{eq:Z1}
\end{equation}
where $\lambda_i$ are the eigenvalues of the transfer matrix $M$. Taking
$\lambda_1$ to be the maximum eigenvalue, we have that,
\begin{equation}
 Z = \lambda_1^P \left[ 1 + \left(\frac{\lambda_2}{\lambda_1}\right)^P 
+ \left(\frac{\lambda_3}{\lambda_1}\right)^P + \ldots \right],
\label{eq:Z2}
\end{equation}
and so in the limit $P \rightarrow \infty$ (i.e.~for a semi-infinite
triangular lattice) we have that $Z = \lambda_1^P$. The free energy is
found via $F = -T \ln Z = -T \ln(\lambda_1^P) = -T P \ln(\lambda_1)$.
Since the total number of sites is $N_{tot} = 2P \times N$, the free
energy per site is given by 
\begin{equation}
f = \frac{F}{N_{tot}} = \frac{-T \ln(\lambda_1)}{2N}.
\label{eq:F}
\end{equation}

With this method, we also investigate the influence of disorder in $J_1$ and
$J_2$. To obtain the
partition function when disorder is present, we instead take the trace
of the product of many transfer matrices, each one using a different set
of values for $J_1$ and $J_2$.  We show results
for which these parameters are chosen from a uniform distribution and also a Gaussian distribution.  

\section{III. Results and Discussion}

\subsection{A. Results in zero field with no disorder}

The transfer matrix method outlined above was used to obtain $S(T)$ and
$C(T)$ results for the TIAF in the absence of a magnetic field, with no
disorder. To investigate the effect of the NNN coupling, we calculated
$S(T)$ and $C(T)$ curves for a range of $J_2$ values including $J_2=0$, i.e. NN
interactions only. Fig.~1 shows entropy per site as a function of
temperature for various antiferromagnetic NNN interaction strengths:
$J_2=0, -0.01, -0.02, -0.05, -0.10$ and $-0.25$ (with $J_1=-1$). As expected, for $J_2=0$
(black curve) we observe a non-zero residual entropy as the
temperature tends to zero, since frustration in the triangular lattice
produces a degenerate ground state when only nearest neighbor
interactions are present. The presence of any non-zero next-nearest
neighbor interaction removes the ground state degeneracy, giving an
entropy which tends to zero at low temperature. For small values of
$J_2$
(e.g.~$J_2=-0.01$), a plateau in the $S(T)$ curve is observed at the value
of the residual entropy for the $J_2=0$ case, before sharply dropping to
$S=0$ as the temperature reaches zero. As the magnitude of $J_2$ increases,
the entropy plateau is gradually rounded, until there is no longer a
plateau visible in $S(T)$ (i.e., for $|J_2| \geq 0.05$). 

Finite size effects can be seen in the inset of Fig.~1. The entropy function can not
have a jump in a finite system, instead that change in entropy will happen over a range of temperatures.
The Monte Carlo study of Rastelli et al \cite{rastelli} allows us to locate the amount of the entropy jump at the
transition and the transition temperature for $J_2=-0.1$. 
These are indicated in the inset figure by a dashed curve. 
Previous Monte Carlo data \cite{rastelli} are only available for a magnitude of $J_2$
greater than or equal to $0.1$, which pushes the first order transition temperature outside
the plateau region of the nearest-neighbor model. Thus, we have developed further
Monte Carlo simulations for $J_2=-0.05$ and $J_2=-0.02$ (and also verified the results for $J_2=-0.1$ \cite{rastelli})
to study the entropy jump at a temperature
in the plateau region of the nearest-neighbor model. These jumps are also shown
in the inset. The data are consistent with the absence of an entropy plateau for $|J_2| \geq 0.05$. For $J_2=-0.02$, we calculate an entropy jump which is consistent with the sharp change in the entropy function observed below the plateau.

\begin{figure}[h]
\begin{center}
 \includegraphics[width=\columnwidth]{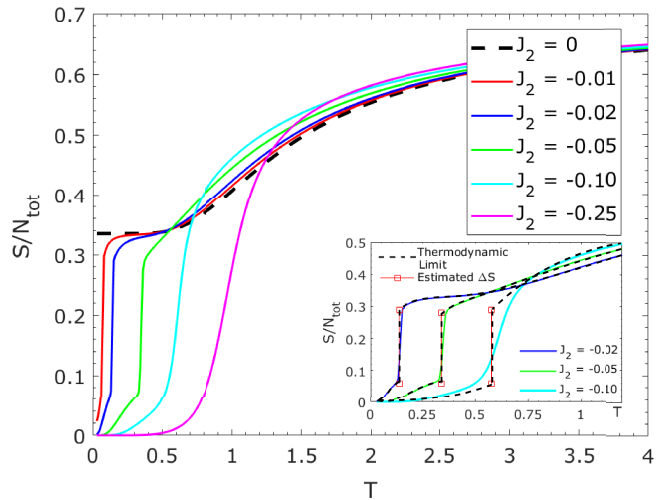}
 \caption{\label{fig1} 
Entropy per site as a function of temperature for the semi-infinite TIAF
geometry, calculated using the transfer matrix method. $S(T)$ curves are
shown for six different values of the NNN interaction $J_2$, with $J_1=-1$ fixed. At $J_2=0$ a
residual ground state entropy is observed at zero temperature.
The inset shows a comparison of our $J_2=-0.1, -0.05$ and $-0.02$ entropy functions with the
entropy jump expected in the thermodynamic limit. The magnitude of the jump and the
transition temperatures are obtained from the hysterisis of the energy function in
the Monte Carlo simulations of up to $96\times 96$ systems.
}
\end{center}
\end{figure}

The specific heat per site is found using the relation
$C=T\frac{\partial S}{\partial T}$, and is shown in Fig.~2 for a range
of $J_2$ values. For non-zero values of $J_2$, a peak in the specific heat is
observed at the temperature where $S(T)$ sharply drops, indicating a
transition to an ordered ground state. The $J_2=0$ specific heat curve
(shown in black) has no peak, since ordering to a non-degenerate ground
state does not occur. As the magnitude of $J_2$ increases, the
peaks in the specific heat are shifted to higher
temperature, consistent with Fig.~\ref{fig1}.

The entropy and specific heat of the infinite triangular lattice
(i.e.~in the thermodynamic limit) were calculated exactly by Wannier
\cite{wannier1950, wannier1973}. An exact expression for $C(T)$ (with NN
interactions only) for the TIAF is given in
\cite{metcalf_thesis}, which is plotted in Fig.~3. The transfer matrix
result for $C(T)$ for our semi-infinite $6 \times \infty$ system is
plotted for comparison, indicating our results are in good agreement
with the exact case in the thermodynamic limit. By integrating the
specific heat we also obtain the exact form of $S(T)$ in the thermodynamic
limit, which is shown in Fig.~4. The numerical result for $S(T)$ for our
semi-infinite geometry is also shown, and the agreement between the numerical and exact results is even closer than it is for specific heat. The Wannier value of the residual
entropy in the thermodynamic limit is $S(0) \approx 0.32306$
\cite{wannier1973}, and for our semi-infinite $6 \times \infty$ system
we obtain $S(0) \approx 0.3350$. 

\begin{figure}[h]
\begin{center}
 \includegraphics[width=\columnwidth]{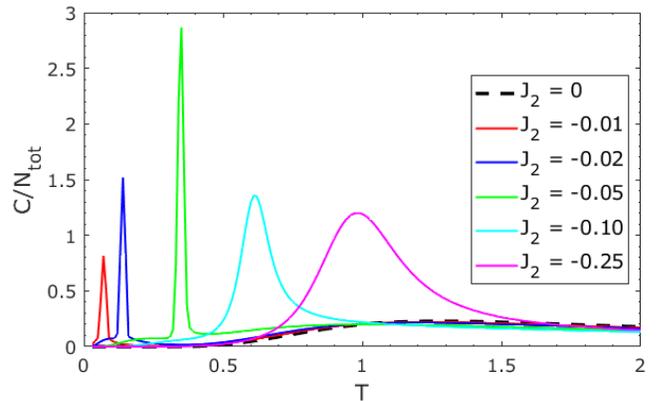}
 \caption{\label{fig2} 
Specific heat as a function of temperature for six different values of
$J_2$, obtained from our $S(T)$ calculation. Peaks in the specific heat occur
at temperatures at which the corresponding $S(T)$ curve sharply drops to
zero.  } 
\end{center} 
\end{figure}

\begin{figure}[h]
\begin{center}
 \includegraphics[width=\columnwidth]{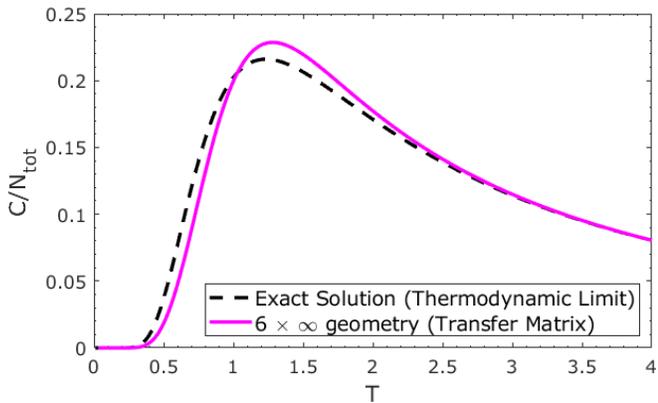}
 \caption{\label{fig3} 
Comparison of $C(T)$ calculated using the transfer matrix method for our semi-infinite 
$6\times\infty$ lattice with the exact result for the TIAF in the thermodynamic limit.
}
\end{center}
\end{figure}

\begin{figure}[h]
\begin{center}
 \includegraphics[width=\columnwidth]{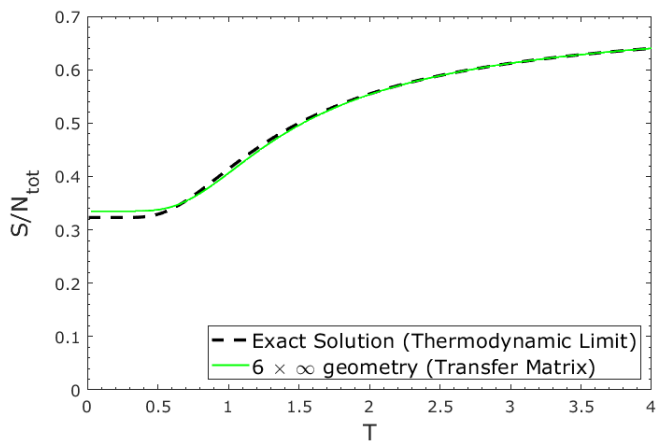}
 \caption{\label{fig4} 
Comparison of $S(T)$ with the exact result in the thermodynamic limit. We
obtain a residual entropy of $S(0) \approx 0.3350$ for our semi-infinite
$6\times\infty$ lattice, which is slightly greater than the exact value
$S(0) \approx 0.32306$ in the thermodynamic limit.
}
\end{center}
\end{figure}

\subsection{B. Results in zero field with Gaussian disorder}

We observe in Fig.~1 that in the absence of NNN interactions, $S(T)$ tends towards the residual entropy value as the temperature is reduced to zero, with a short plateau appearing at low temperature. With a non-zero $J_2$, we find that a plateau at the residual entropy value exists for a finite temperature window, before $S(T)$ drops to zero. As the magnitude of $J_2$ increases (i.e.~for $|J_2| \geq 0.05$) this plateau weakens and we observe the $S(T)$ curve smoothly decreasing to zero. In order to determine the robustness of such entropy plateaus, we now consider the effect of Gaussian disorder (in the nearest-neighbor variable $J_1$) on the form of the $S(T)$ curve. Fig.~5(a) shows the effect of increasing levels of Gaussian disorder in $J_1$ in the absence of next-nearest-neighbor interactions. To obtain $S(T)$ values with disorder, the trace of the product of 101 transfer matrices was taken, each one containing $J_1$ values drawn from a Gaussian distribution for each occurrence, i.e., each individual NN coupling in the lattice has a randomly chosen interaction strength. The particular set of $J_1$ values used was stored and used for each temperature increment. 

We label Gaussian distributions by $G(\mu, \sigma)$ where $\mu$ and $\sigma$ denote the mean and standard deviation respectively. As $\sigma$ is increased, the plateau at the value of the residual entropy is gradually weakened, and we eventually observe the $S(T)$ curve approaching zero with no plateau. At all temperatures, the entropy per site is lower for increasing levels of disorder, and we also find that for relatively low levels of disorder (e.g.~$J_1=G(-1,0.02)$) an entropy plateau persists to quite low temperatures ($T\approx 0.2$). A logarithmic temperature scale emphasizes the influence of $\sigma$ on the form of the entropy plateau at low values of $T$, as illustrated in Fig.~5(b). For wider distributions (i.e., $\sigma \geq 0.05$), a short plateau is no longer observed. Even in the absence of a NNN interaction, we see that the introduction of any amount of disorder in $J_1$ leads to a non-degenerate ground state with $S(T)$ approaching zero at $T=0$. Moreover, the presence of weak disorder, i.e.~with a standard deviation of just a few percent of the mean $J_1$ value, is enough to remove any sign of a plateau at low temperatures. This suggests that for the TIAF system, the existence of entropy plateaus is highly sensitive to disorder in the nearest-neighbor interaction.

\begin{figure}[h]
\begin{center}
 \includegraphics[width=\columnwidth]{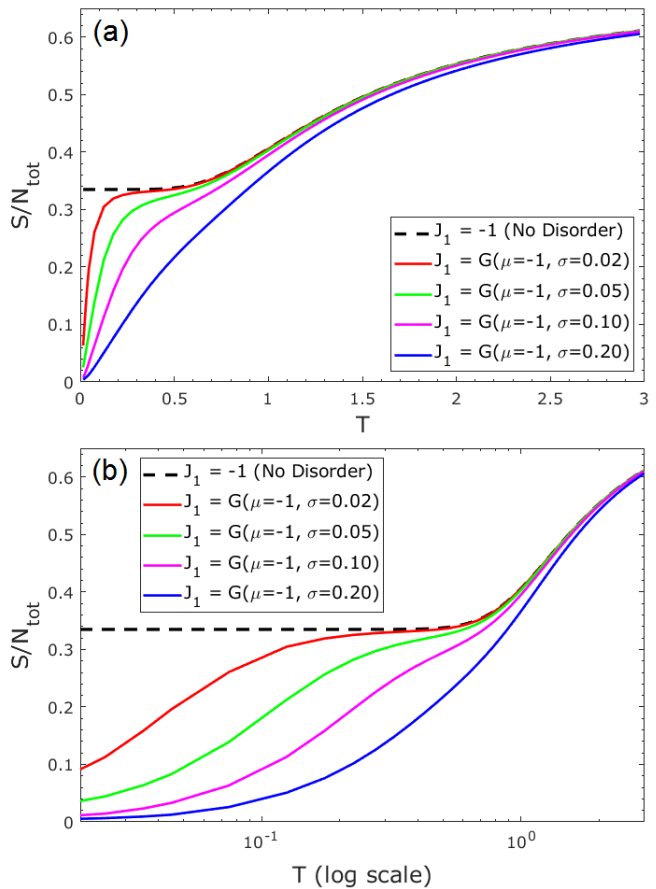}
 \caption{\label{fig5} 
(a) $S(T)$ results with Gaussian disorder in $J_1$ are shown for various values of $\sigma$, with $J_2=0$. The mean of the distribution is fixed at $\mu = -1$ in each case. The entropy curve in the absence of disorder is shown in black for comparison. (b) The same $S(T)$ results as above shown on a logarithmic temperature scale,  emphasizing differences in plateau rounding at low $T$.
}
\end{center}
\end{figure}

\subsection{C. Results in magnetic field with no disorder}

Introducing a magnetic field aligned parallel to the Ising axis,
the Hamiltonian given by Eq.~(\ref{eq:Ham}) now has a non-zero value of $B$ in
the final term. Using the same transfer matrix procedure with this
Hamiltonian, we again obtained $S(T)$ and $C(T)$ plots at different values
of $B$, for various values of $J_2$. Fig.~6 shows $S(T)$ for three 
NNN interaction strengths: (a) $J_2=0$, (b) $J_2=-0.01$ and (c)
$J_2=-0.10$. For the $J_2=0$ case, we again find that for $B=0$, we have a non-zero
residual entropy as the temperature tends to zero. As noted in
\cite{metcalf1978}, a critical field value exists for antiferromagnetic
Ising lattices at $B_c = z|J_1|$, at which there is degeneracy in the ground state. 
Here $z=6$ for the triangular lattice,
and we take $|J_1|=1$.
Hence we observe a non-zero residual entropy again at $B=6$. For
all other magnetic field values, the entropy tends to zero at low
temperature since the ground state degeneracy due to frustration is
removed.

\begin{figure}[h]
\begin{center}
 \includegraphics[width=\columnwidth]{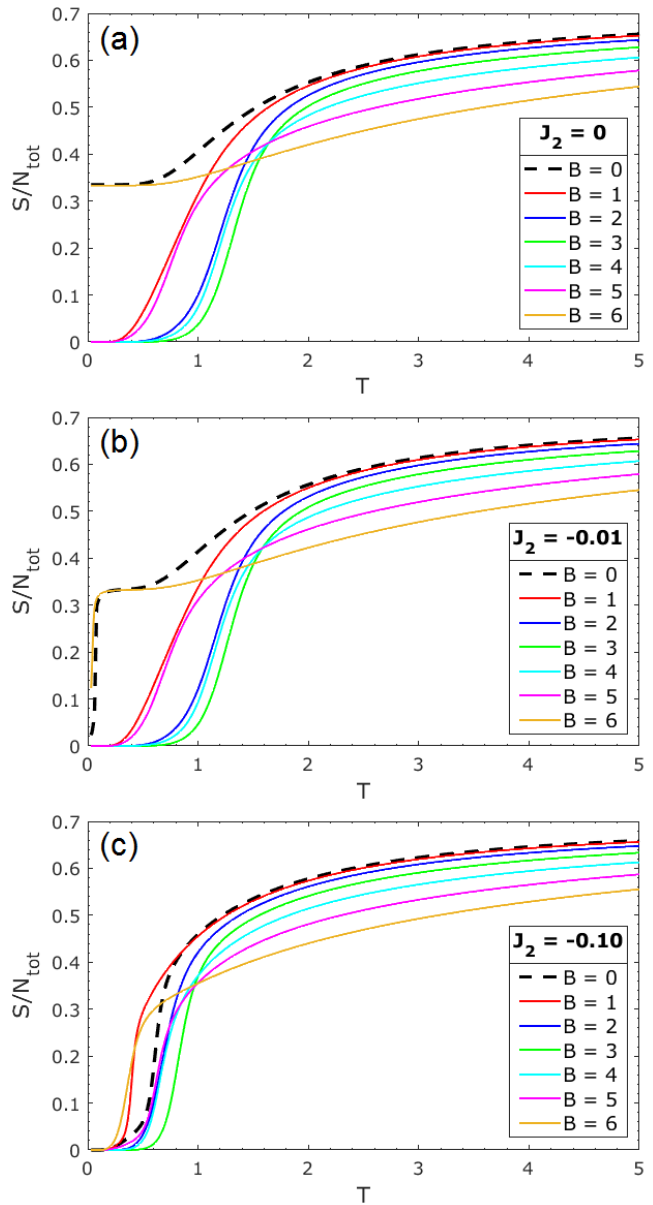}
 \caption{\label{fig6} 
$S(T)$ results in the presence of a magnetic field, with no disorder.
Field strengths ranging from $B=0$ to the TIAF critical field value
$B_c=6$ are shown. The magnitude of the NN interaction strength $J_1$ is set
to 1. Entropy curves are shown for three different $J_2$ values: (a)
$J_2=0$, (b) $J_2=-0.01$, and (c) $J_2=-0.10$.  
}
\end{center}
\end{figure}

\begin{figure}[h]
\begin{center}
 \includegraphics[width=\columnwidth]{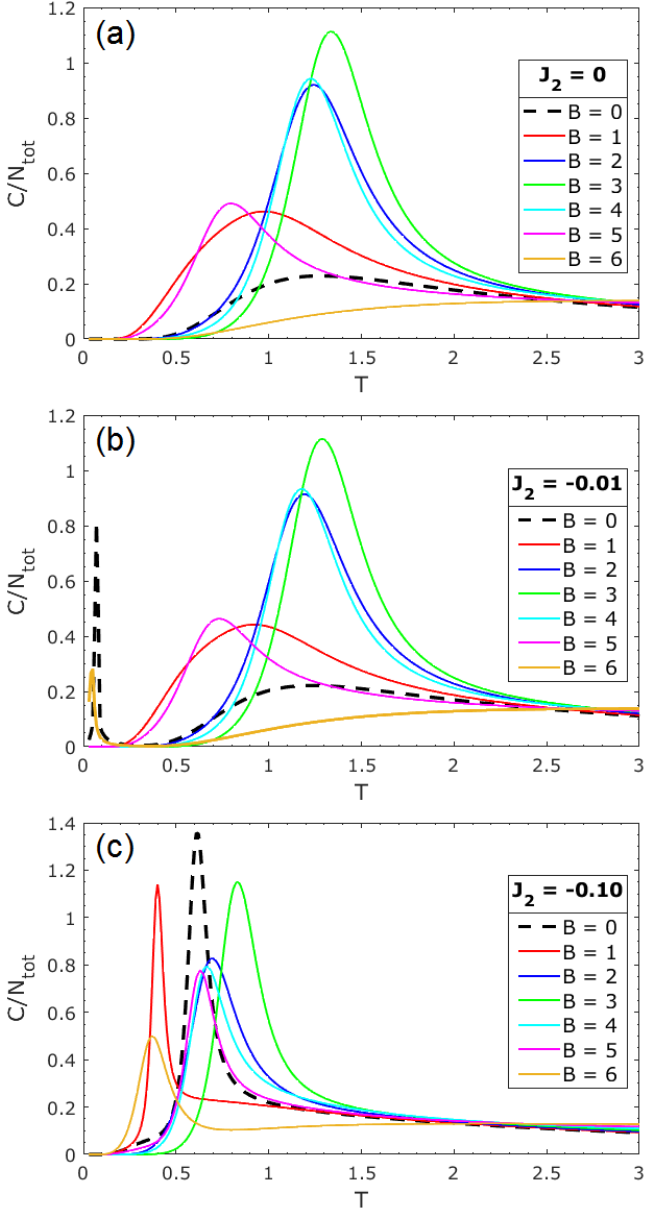}
 \caption{\label{fig7} 
$C(T)$ results in the presence of a magnetic field, with no disorder.
Specific heat curves are shown for three different $J_2$ values: (a)
$J_2=0$, (b) $J_2=-0.01$, and (c) $J_2=-0.10$. Low temperature peaks in
$C(T)$ are observed for both $B=0$ and $B=6$ when there is a small NNN 
interaction present, i.e.~for $J_2=-0.01$.
}
\end{center}
\end{figure}

When a NNN interaction is introduced, as in Fig.~6(b), where $J_2=-0.01$,
there is no residual entropy even at $B=0$ or $B=6$, since the ground-state degeneracy is removed. For small values of $J_2$ we observe a
rounded plateau in $S(T)$ for $B=0$ and $B=6$. As the magnitude of $J_2$ increases, we no longer observe a plateau, and
the entropy per site smoothly falls from $\ln 2$ to zero as the
temperature is lowered. As in the previous section, plots of the
specific heat (at various magnetic field values) were obtained for
$J_2=0, -0.01$ and $-0.10$, as shown in Figs.~7(a)--(c). Comparing the
$B=0$ case in Fig.~7(a) and Fig.~7(b), we see that introducing a small
non-zero NNN interaction (i.e.~$J_2=-0.01$) produces a peak in the
specific heat, indicating a transition to a non-degenerate ground state and the absence of residual entropy at $T=0$. Similarly, we also observe a peak in $C(T)$ at low temperature for $B=6$, when $J_2=-0.01$.
Increasing the magnitude of $J_2$ further, we find that the locations of
the peaks are shifted to lower temperature, for all values of $B$ between $B=0$ and $B=6$.

With a non-zero magnetic field, we can obtain free energies at a fixed
temperature for a range of $B$ values and obtain the magnetization (per
site) using $M=-\frac{\partial F}{\partial B}$. $M(B)$ curves were
obtained at $T=0.05$, $T=0.2$ and $T=2$ for three different values of
$J_2$, as shown in Fig.~8: (a) $J_2=0$, (b) $J_2=-0.01$, and (c)
$J_2=-0.10$. We find that at relatively high temperature (i.e.~$T=2$),
magnetization per site increases linearly with magnetic field, and no
plateaus occur. As temperature is lowered, magnetization plateaus are
observed. The plateaus become less rounded and more step-like as the
temperature is lowered further. For $J_2=0$ and $J_2=-0.01$, a single
plateau is observed at $M=1/3$, but for $J_2=-0.1$ we observe another
plateau at $M=1/2$, suggesting the $M=1/2$ plateau phase is only
observed if the NNN interaction is sufficiently strong. Indeed, the $J_2$
dependence of the width of the $M=1/2$ plateau at finite temperature
(shown in Fig.~9) illustrates that a well defined plateau appears only
when $|J_2|$ exceeds some threshold value, with plateau width increasing
approximately linearly with $|J_2|$ thereafter. At a temperature of
$T=0.05$, a plateau at $M=1/2$ is apparent for $|J_2| \geq 0.04$.
Decreasing the magnitude of $J_2$ gradually rounds the plateau until it
is no longer present, and the magnetization per site increases smoothly
from $1/3$ to full saturation.  From Fig.~8(c) we can see that for
$J_2=-0.10$ (at $T=0.2$), we have an $M=0$ stripe phase for approximately
$0<B<1$, an $M=1/3$ plateau phase in the region $1.2<B<4.1$, and an
$M=1/2$ phase for $4.2<B<6$. Greater values of magnetic field produce a
fully spin-polarized phase with $M=1$. We also find that as the
temperature increases, the rounding of the $M=1/2$ plateau is more
pronounced than at $M=1/3$, which at $J_2=-0.10$ and $T=0.2$ remains
step-like, as shown in Fig.~8(c) (red curve).

\begin{figure}[t]
\begin{center}
 \includegraphics[width=\columnwidth]{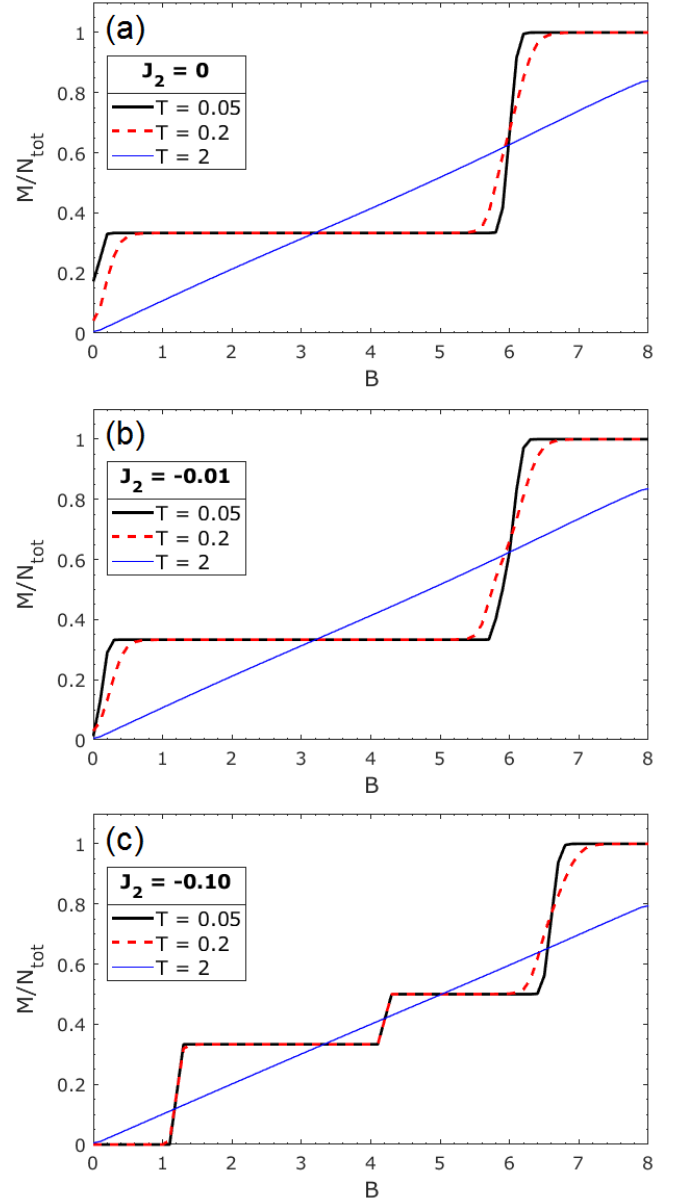}
 \caption{\label{fig8} 
$M(B)$ results at finite temperature for different values of $J_2$: (a)
$J_2=0$, (b) $J_2=-0.01$, and (c) $J_2=-0.10$, without disorder. For
$J_2=-0.10$, step-like magnetization plateaus can be seen at both $M=1/3$
and $M=1/2$. In each plot, $M(B)$ curves for three different temperatures
are shown: $T=0.05$, $T=0.2$, and $T=2$.
}
\end{center}
\end{figure}

\begin{figure}[h]
\begin{center}
 \includegraphics[width=\columnwidth]{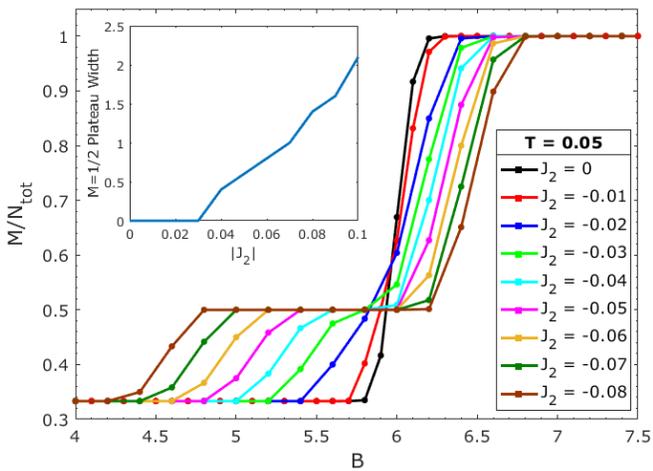}
 \caption{\label{fig9} 
$M(B)$ curves around the $M=1/2$ plateau region are shown for various
values of $J_2$, at a fixed finite temperature of $T=0.05$. The inset graph
shows the dependence of the $M=1/2$ plateau width (in units of $B$) 
upon the magnitude of $J_2$.
}
\end{center}
\end{figure}

\subsection{D. Results in magnetic field with uniform and Gaussian disorder}

Although rounding of the magnetization plateaus in the ideal TIAF system
is illustrated here at finite temperature, at $T=0$ the magnetization
per site increases in discrete steps between $0$, $1/3$, $1/2$ and $1$. However,
in low-temperature measurements of TIAF materials such as TmMgGaO$_4$
\cite{cevallos, li}, distinct plateaus in magnetization are absent,
which has been ascribed to the presence of disorder in inter-site
interactions and coupling to the magnetic field, which weakens or
removes these plateaus entirely. The influence of disorder on $M(B)$ for
the TIAF with both NN and NNN interactions has been studied previously
for a finite $6 \times 6$ cluster by Li et al \cite{li}, where it was
found that introducing disorder produced magnetization curves in
agreement with experiment. Motivated by this work, we studied the
influence of disorder strength on the form of the magnetization
plateaus, and also investigated the relative importance of disorder in
$J_1$ and $J_2$.

Using the transfer matrix approach to obtain $M(B)$, one introduces
disorder in $J_1$ and $J_2$ by generating random values of these parameters
from a chosen distribution. For
a given set of $J_1$ and $J_2$ values (for all individual NN and NNN couplings in the lattice) we produce the corresponding
transfer matrices as before, however the partition function is now
obtained by taking the trace of the product of $P$ transfer matrices, each
one containing different parameter values. We use $P = 101$
and parameter values drawn from both a uniform distribution and a Gaussian distribution in the results
presented here, with temperature fixed at $T=0.2$. Uniform distributions are denoted by $U(J_{min},J_{max})$ where $J_{min}$ and $J_{max}$ are the boundaries of the distribution, which has a width $J_{max}-J_{min}$. As shown in Fig.~10(a), we observe that strong plateaus at both $M=1/3$ and $M=1/2$ remain for $J_1=U(-1.2,-0.8)$ and $J_2=U(-0.16,-0.04)$, i.e.~uniform distributions with mean values of $J_1=-1$ and $J_2=-0.1$. As the distribution width is increased, the plateaus are rounded further, and we find that both the $M=1/2$ and $M=1/3$ plateaus are eventually no longer observable, e.g.~for $J_1=U(-1.4,-0.6)$ and $J_2=U(-0.2,0)$. There is an indication that the $M=1/3$ plateau may be more robust to disorder than the $M=1/2$ plateau, since as the level of disorder increases, the plateau at $M=1/2$ is lifted while a  short plateau remains observable at $M=1/3$, which can be seen for $J_1=U(-1.3,-0.7)$ and $J_2=U(-0.18,-0.02)$. When the disorder strength is increased further (i.e.~to $J_1=U(-1.4,-0.6)$ and $J_2=U(-0.2,0)$),
the weak plateau at $M=1/3$ is no longer present,
and one obtains a magnetization curve quite similar to the recent experimental result for TmMgGaO$_4$ \cite{li}.

We also investigated introducing disorder in only one of the parameters $J_1$ or $J_2$ as shown in Fig.~10(b), using uniform distributions $J_1=U(-1.4,-0.6)$ and $J_2=U(-0.2,0)$, again with mean values $J_1=-1$ and $J_2=-0.1$. We find that with disorder in $J_2$ only, both plateaus at $M=1/3$ and $M=1/2$ are
present, and the magnetization curve remains similar to the zero-disorder case. With disorder in $J_1$ only, both plateaus are completely removed, and $M(B)$ is essentially identical to our result with disorder in both $J_1$ and $J_2$ combined. This suggests that disorder in $J_1$ only is sufficient to eliminate both magnetization plateaus, giving an $M(B)$ curve similar to experiment, provided $|J_1|$ exceeds $|J_2|$ by approximately an order of magnitude, as in this study. In this case, magnetization plateaus in the TIAF system are robust to disorder solely in $J_2$, even when the width of the parameter distribution spans $\pm 100\%$ of the mean value, i.e.~for $J_2=U(-0.2,0)$.  With disorder in $J_2$ only, step-like transitions between magnetization plateaus are still present, and there is only a slight rounding of the magnetization curve compared to the zero-disorder case, even for $J_2=U(-0.2,0)$. We also find that disorder in the magnetic field is relatively insignificant, and one can obtain a result qualitatively similar to experiment with disorder in $J_1$ and $J_2$ only. 

The form of the $M(B)$ curves in the presence of Gaussian disorder in $J_1$ and $J_2$ (instead of uniform disorder) was also investigated, as shown in Fig.~11. The Gaussian distributions for $J_1$ and $J_2$ were chosen to have mean values of -1 and -0.1 respectively, with the standard deviation of the distribution of $J_2$ values fixed at $\sigma=0.1/\sqrt{3}$. The width of the distribution of $J_1$ values was varied and $M(B)$ results compared to the zero-disorder case. Three Gaussian distributions were used, which were chosen to have same standard deviations for $J_1$ as the three uniform distributions shown in Fig.~10(a), where we have used $\sigma=(b-a)/\sqrt{12}$ for any uniform distribution $U(a,b)$. As with uniform disorder, increasing the width of the distribution gradually weakens the magnetization plateaus at $M=1/3$ and $M=1/2$ until both are no longer visible, which occurs when the standard deviation of $J_1$ values approaches $\sigma=0.4/\sqrt{3}$. The strong similarity between our results for uniform and Gaussian disorder indicates that the exact form of the distribution used is relatively unimportant in determining the robustness of magnetization plateaus in the TIAF system. 

\begin{figure}[h]
\begin{center}
 \includegraphics[width=\columnwidth]{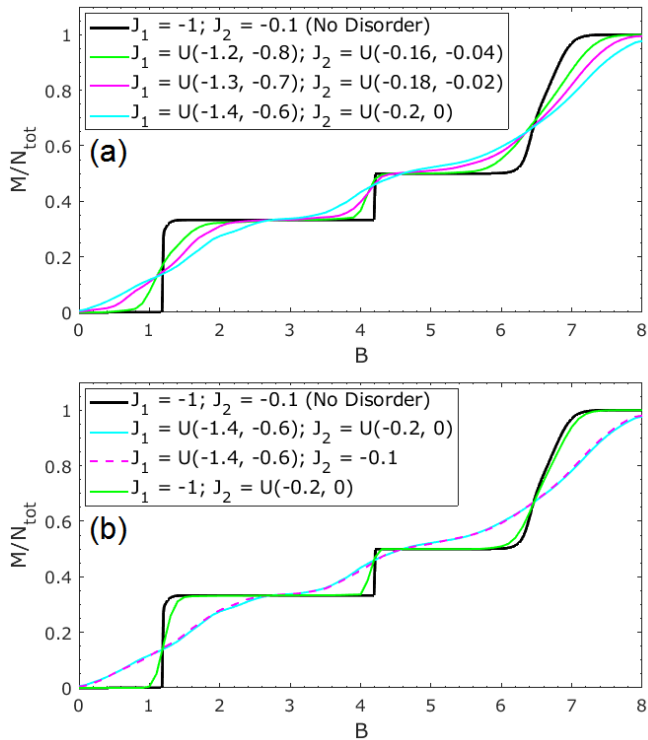}
 \caption{\label{fig10} 
(a) $M(B)$ results for three different levels of uniform disorder, in both $J_1$ and $J_2$ combined, with the zero-disorder result shown in black for comparison. In each case, the uniform distributions of $J_1$ and $J_2$ used are centered at -1 and -0.1 respectively. (b) Additional $M(B)$ results are shown for uniform disorder in $J_1$ only ($J_1=U(-1.4, -0.6)$ with $J_2=-0.1$) and in $J_2$ only ($J_2=U(-0.2,0)$ with $J_1=-1$).
}
\end{center}
\end{figure}

\begin{figure}[h]
\begin{center}
 \includegraphics[width=\columnwidth]{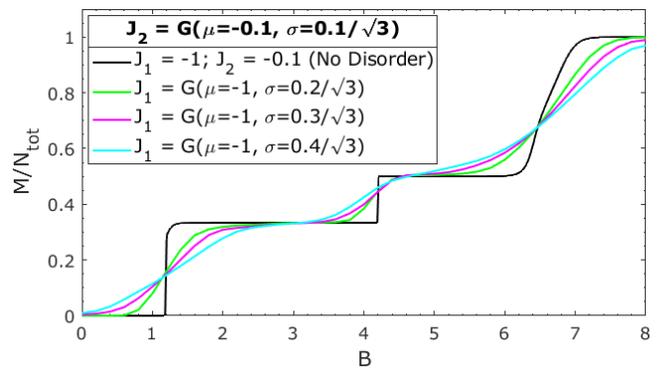}
 \caption{\label{fig11} 
$M(B)$ results for three different levels of Gaussian disorder, in both $J_1$ and $J_2$ combined, with the zero-disorder result shown in black for comparison. The standard deviations of the distributions of $J_1$ values match those used in Fig.~10(a) for the case of uniform disorder.
}
\end{center}
\end{figure}

\section{IV. Summary and Conclusions}

In this work we have studied the rounding and in some cases complete absence of entropy and magnetization
plateaus for the triangular lattice antiferromagnet, with both NN and
NNN interactions in a magnetic field, with and without quenched disorder.
In particular, we have found that in the ideal TIAF, increasing $|J_2|$
tends to round and quickly remove the plateau in $S(T)$ near
the theoretical residual entropy value at low temperature. The plateau-like feature
is replaced by a sharp drop in entropy at the first-order transition. The strength
of the second-nearest-neighbor interaction also determines if a
magnetization plateau at $M=1/2$ is present at finite temperature, and
controls the width of the plateau. For sufficiently large $J_2$, a distinct
plateau at $M=1/2$ will be visible, which is gradually rounded as the
magnitude of $J_2$ is lowered, until a plateau no longer remains. 

In order to model realistic TIAF materials such as TmMgGaO$_4$, we
studied the influence of disorder in $J_1$ and $J_2$ on the form of the
entropy and magnetization curves. 
We find that with nearest-neighbor interactions alone, rounded entropy plateaus are quite sensitive to disorder in the exchange variable, and they are no longer observed when the width of the $J_1$ distribution exceeds $\sigma \approx 0.05 J_1$. For weaker levels of disorder, a plateau at the residual entropy value persists to low temperatures (around $T=0.2$ for $\sigma = 0.02 J_1$). Consequently, we expect rounded entropy plateaus to be observable in TIAF systems at low temperatures only if 
the second neighbor interactions are less than a few percent and 
there is a significant absence of quenched disorder in the system.
Our $M(B)$ results with disorder are close to recent experimental observations \cite{li},
confirming the presence of second-neighbor interactions and disorder in the system.

More generally, we conclude that the existence of well-defined entropy plateaus requires a fair amount of fine-tuning of the system,
so whether they will be observed in a generic frustrated magnet is unclear.
The spin-ice system is clearly special. 
The fact that a residual entropy plateau is seen in model simulations with 
arbitrary strength long-range dipolar interactions in addition to 
nearest-neighbor exchange interactions \cite{gingras,moessner,ice-review} shows their robustness. 
One might have expected these long-range interactions to remove the ground state degeneracy, 
and the corresponding zero-point entropy. But, it has been shown that a `model dipole' interaction can be constructed that has exactly the same ground states as the nearest-neighbor model \cite{model-dipole, model-dipole2}. Remarkably, the dipolar interaction on the pyrochlore lattice has the noteworthy property of differing only slightly from this model interaction, and at short distances only. This robustness is presumably a manifestation of the emergent gauge theory.

Independent of the issue of fine-tuning, there are strong experimental challenges in looking for these entropy 
plateaus in real materials. The need to have clean low-disorder material and to be able to isolate the magnetic
contribution to heat capacity and entropy from phonons and other degrees of freedom can be formidable.
We hope our work will motivate further work on entropy plateaus in
frustrated magnets and also in strongly-correlated electron systems, where 
a residual entropy phase may be a precursor to intertwined and competing orders \cite{FKT}.

\section{Acknowledgements}
The authors thank W. Pickett for useful discussions. This work is supported in part by DOE grants DE-SC0014671 (RTS) and DE-FG02-04ER46111 (OB and CF) and by the U.S. National Science Foundation DMR grant number 1855111 (RRPS).

\end{document}